\documentclass{JAIS}

\journal{JAIS-ID}
\vol{2022}

\received{xx January 2022}
\published{xx March 2022}

\def\be{\begin{equation}}
\def\ee{\end{equation}}
\def\bea{\begin{eqnarray}}
\def\eea{\end{eqnarray}}

\usepackage{caption}
\usepackage{lineno}
\usepackage[colorlinks=true]{hyperref}

\begin{document}

\title{Muography in Colombia: simulation framework, instrumentation and data analysis}

\author{J. Peña-Rodríguez\auno{1}\auno{*}, A. Vesga-Ramírez\auno{2}, A. Vásquez-Ramírez\auno{1}, M. Suárez-Durán\auno{3}, R. de León-Barrios\auno{1}, D. Sierra-Porta\auno{4}, R. Calderón-Ardila\auno{5}, J. Pisco-Guavabe\auno{1}, H. Asorey\auno{5,6}, J. D. Sanabria-Gómez\auno{1},\\ and L. A. Núñez\auno{1,7}}
\address{$^1$Escuela de Física, Universidad Industrial de Santander, Bucaramanga, Colombia}
\address{$^2$International Center for Earth Sciences, Comisión Nacional de Energía Atómica, Buenos Aires, Argentina}
\address{$^3$Université Libre de Bruselles, Bruselles, Belgium}
\address{$^4$ Departamento de Ciencias Básicas, Universidad Tecnológica de Bolivar, Cartagena de Indias, Colombia.}
\address{$^5$Instituto de Tecnologías en Detección y Astropartículas, Centro Atómico Constituyentes, Buenos Aires, Argentina}
\address{$^6$Consejo Nacional de Investigaciones Científicas y Técnicas, Buenos Aires, Argentina.}
\address{$^7$Departamento de F\'{\i}sica, Facultad de Ciencias, Universidad de Los Andes, M\'{e}rida, Venezuela}
\address{$*$jesus.pena@correo.uis.edu.co}

\begin{abstract}
We present the Colombo-Argentinian Muography Program for studying inland Latin-American volcanoes. It describes the implementation of a simulation framework covering various factors with different spatial and time scales: the geomagnetic effects at a particular geographic point, the development of extensive air showers in the atmosphere, the propagation through the scanned structure and the detector response. Next, we sketch the criteria adopted for designing, building, and commissioning {\it MuTe}: a hybrid Muon Telescope based on a composite detection technique. It combines a hodoscope for particle tracking and a water Cherenkov detector to enhance the muon-to-background-signal separation due to extensive air showers' soft and multiple-particle components. {\it MuTe} also discriminates inverse-trajectory and low-momentum muons by using a picosecond Time-of-Flight system. We also characterise the instrument's structural –mechanical and thermal– behaviour, discussing preliminary results from the background composition and the telescope-health monitoring variables. Finally, we discuss the implementations of an optimisation algorithm to improve the volcano internal density distribution estimation and machine learning techniques for background rejection.
\end{abstract}

\maketitle


\section{Volcanoes in Latin-America and Colombia}
Latin America is part of the Pacific Ring of Fire, with more than 200 Holocene volcanoes sculpting the Andean mountain range traversing the continent. Almost 20 million people live close to active volcanic zones ($<$ 100 km). Four geographic areas group volcanoes in South America: Northern, Central, Southern, and Austral, with more than 600 eruptions reported since 1532, causing 25000 casualties. The most deadly was the disaster that occurred with the lahars of the Nevado del Ruiz  (23000 deaths), Colombia, in 1985 \cite{Stern2004}. 

Volcanoes cause ash clouds, eruptions, pyroclastic flows, lahars, debris avalanches, and earthquakes. The previous monitoring of related phenomena allows the prevention of disasters and casualties. Information on seisms, ground deformation, gas emissions, and thermal anomalies helps to alert authorities and citizens about possible hazards.  About 60$\%$ of Andean volcanoes remain unmonitored due to lack of resources, remote access and dangerous conditions for the  installations of equipment locally \cite{Reath2019}.

Colombia has more than a dozen active volcanoes (Azufral, Cerro Negro, Chiles, Cumbal, Doña Juana, Galeras, Machín, Nevado del Huila, Nevado del Ruíz, Nevado Santa Isabel, Nevado del Tolima, Puracé, and Sotará) clustered in three main groups along the Cordillera Central, the highest of the three branches of the Colombian Andes. Most of these volcanoes represent a significant risk to the nearby population in towns and/or cities \cite{cortes2016, agudelo2016, munoz2017}.

The Galeras, Cerro Machín, and Nevado del Ruíz volcanoes represent a high risk for Colombian citizens. Evidence of that was the Nevado del Ruiz eruption on November 13, 1985, which involved the partial melting of the glacier cap and consequent lahars. Mood flows reached the municipality of Armero-Tolima, causing more than 23000 causalties \cite{VesgaRamirez2020}.

The Cerro Machín volcano, with a crater of $2.4$~km diameter and $450$~m height, could cause a future pyroclastic eruptive episode reaching an area of $10$~km$^2$ around the volcano edifice \cite{Murcia2010}. The similarities in morphologies between two volcanic must be considered in the risk evaluation of an eventual eruption. The Machín\footnote{\url{https://es.wikipedia.org/wiki/Cerro_Machin}} volcano has a morphology  similar to the Chichón/Chichonal\footnote{\url{https://en.wikipedia.org/wiki/Chichonal}} volcano in southern state of Chiapas in México \cite{VesgaRamirez2020}. The Chichón has been the most deadly eruption in Mexico, creating an ash cloud covering $200$~km, $\sim2000$ causalities, and the displacement of $\sim20000$ citizens in March 1982. \cite{DelaCruzReyna2009}.

\section{Muography in the Latin American region}

Muography is a non-invasive technique for scanning the inner structure of volcanoes. This method uses the attenuation of atmospheric muons after crossing the volcanic edifice.  The emerging muon flux measurements at different directions give estimations for the density distributions of  volcanic edifices \cite{VesgaRamirez2020}.

\begin{figure}[h!]
\centering
\includegraphics[height=3.25in]{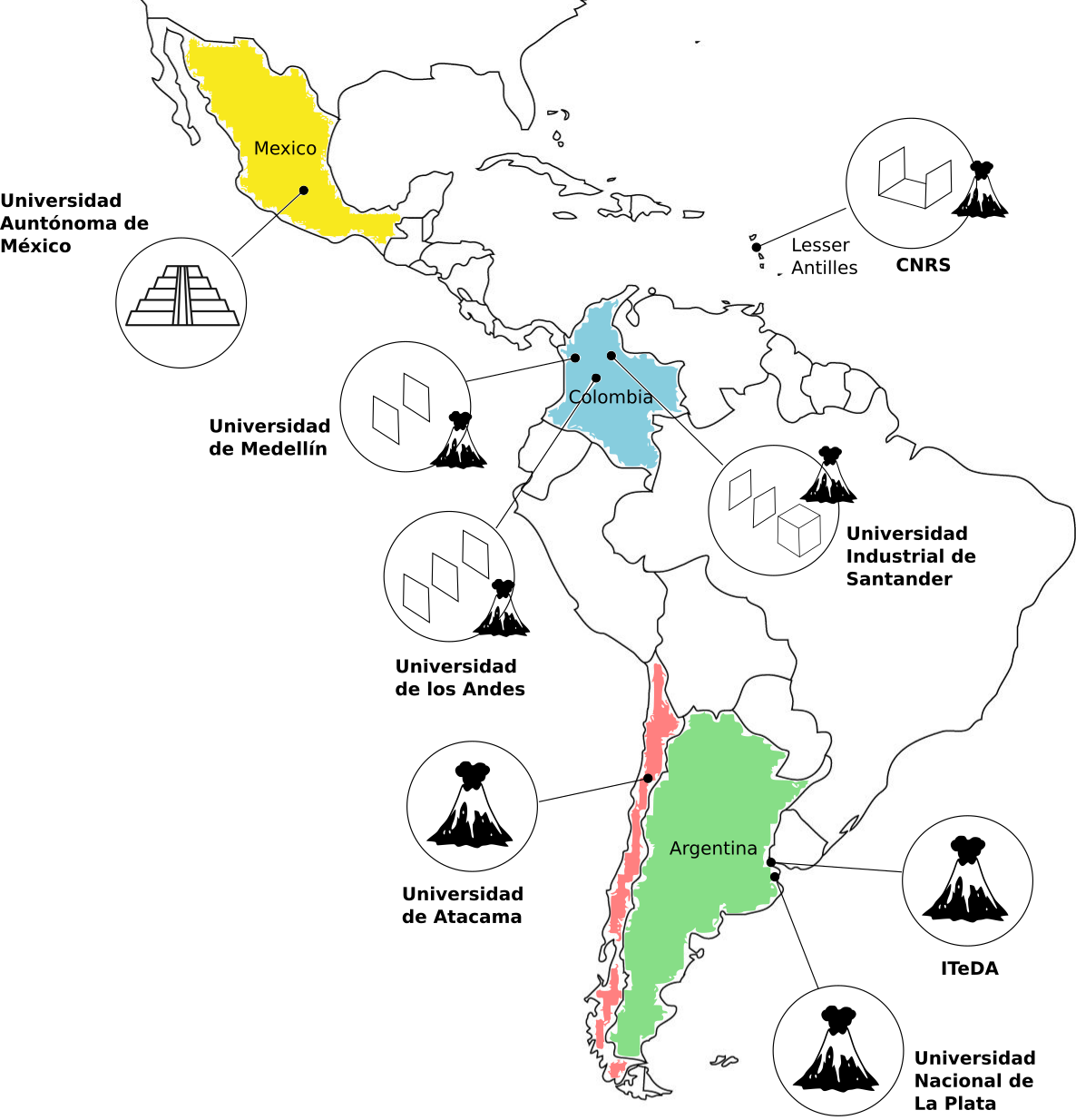}
\caption{Geographical distribution of Universities/Institutions actively working in muography in Latin America. Most of them are focused on muography applied to vulcanology. Argentina and Colombia register most activities with groups cooperating in different areas.}
\label{f:muography_LA}
\end{figure}
Several Latin American Universities/Institutions (see Figure \ref{f:muography_LA}) work in muography due to the number of active volcanoes along the Andean mountain range and the risk they represent. The Latin American muography consists in three research areas: creation of simulation frameworks, development of instrumentation and implementation of data analysis methods \cite{SierraportaEtal2018, PenarodriguezEtal2018B}.

Muography simulation tools \& frameworks estimate the atmospheric muon flux \cite{Asorey2018,GrisalescasadiegosSarmientocanoNunez2020}, muon propagation through the geological target \cite{VesgaRamirez2020, Tapia2017,  Parra2019}, detector response \cite{VsquezRamrez2020, Guerrero2019, MenchacaRocha2014} and density optimisation \cite{VesgaRamrez2021} by using Monte Carlo codes. Development of muography instruments have been addressed from different perspectives. The use of continuous scintillators \cite{Guerrero2019},  scintillators strips \cite{Lesparre2012}, hybrid muon telescopes for enhancing the muon-to-background ratio \cite{PeaRodrguez2020, PeaRodrguez2019}, and development of new technologies \cite{CaldernArdila2020}. 

The Muon Telescope implements machine learning algorithms for particle identification of signal/background using the deposited energy, time-of-flight, and momentum \cite{PeaRodrguez2021}. Muography brings data analysis challenges from raw data treatment to solving the density estimation inverse problem. Simulated annealing and Bayesian dual inversion algorithms tackle the inverse problem to infer the target density distribution \cite{VesgaRamrez2021, Lesparre2011}.

\section{The Muon Telescope}

The Muon Telescope (MuTe) project involves designing and implementing a muon detector for scanning volcanoes. The project boosts the muography research area in Colombia in close collaboration with Argentinian colleagues, providing an innovation ecosystem between the two countries. This R \& D area identifies three lines of action: the simulation framework, the instrumentation and online/offline data analysis. The simulation framework implements Monte Carlo simulations to estimate the muon flux at any geographical point and then estimate the energy loss of the crossing muons and the detector response. The instrumentation branch manages the steps of designing, implementing, and calibrating the muon telescope. The data analysis branch processes the recorded data and develops algorithms for background rejection and density inverse problem-solving.

The MuTe combines two detection techniques. A hodoscope measures the muon flux by tracking the particle trajectories and a water Cherenkov detector (WCD) which records the deposited energy of particles traversing the MuTe. A time-difference system for the hodoscope measures the time-of-flight of particles impinging the detector \cite{PeaRodrguez2020}. The particle identification system of MuTe allows the rejection of muography background sources: the soft component of Extended Air Showers, backscattering particles, and forward scattering muons \cite{PeaRodrguez2021}. 

A photo-voltaic system supplies the MuTe with autonomy for up to 6 days in cloudy conditions. The MuTe daily reports the hodoscope and WCD status: power consumption, average particle flux, detector temperature, atmospheric pressure, and environmental humidity.

\section{{Simulation framework}}

\begin{figure}[h!]
\centering
\includegraphics[height=2.5in]{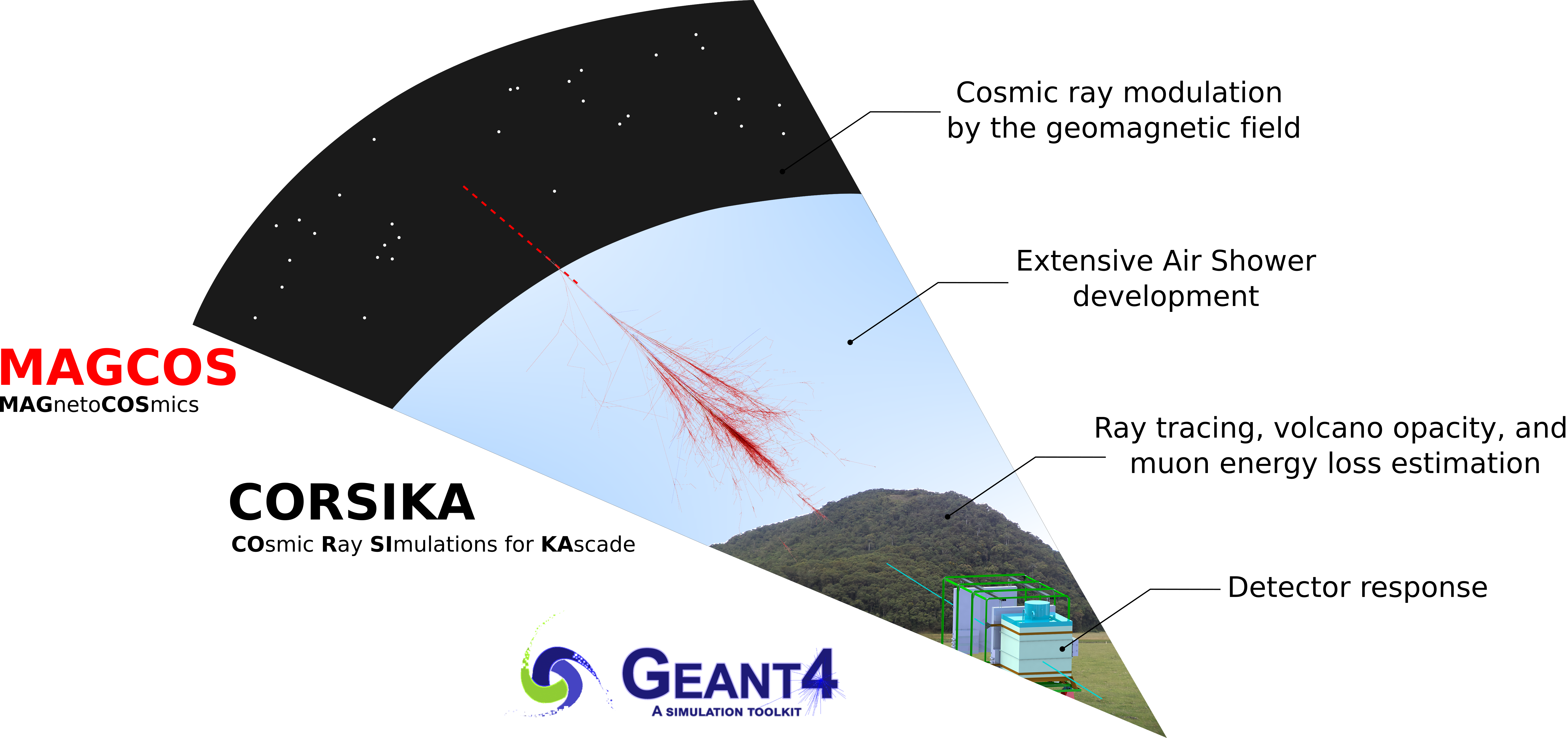}
\caption{Muography simulation framework. MAGCOS computes the cosmic ray modulation due to the geomagnetic field, and CORSIKA estimates the atmospheric particle flux at the observation place. MUSIC, Python, and MATLAB scripts compute the ray tracing, volcano opacity, and emerging muon flux (considering energy loss and scattering). GEANT4 simulates the detector response.}
\label{f:simu_frame}
\end{figure}

As shown in Figure \ref{f:simu_frame}, the simulation scheme for the MuTe project implements the ARTI framework from the Latin-American Giant Observatory collaboration. This framework considers three essential factors with different spatial and time scales: the geomagnetic effects, the development of the extensive air showers in the atmosphere, and the detector response at ground level \cite{VsquezRamrez2020, AsoreyEtal2016a, SarmientocanoEtal2020}. 

In the first step, a MAGNETOCOSMICS\footnote{\url{http://cosray.unibe.ch/~laurent/magnetocosmics/}} based script computes the primary cosmic ray flux impinging the earth atmosphere after modulation due to the geomagnetic field rigidity. Then, a second step uses the  CORSIKA\footnote{\url{https://www.iap.kit.edu/corsika/}} script simulates the interaction of primary cosmic rays and the Earth's atmosphere. We compute the spectrum of atmospheric particles reaching the observation place and the angular (zenith and azimuth) dependency of the muon flux \cite{valencia2016}. We found secondary cosmic rays at Cerro Machín volcano (4$^{\circ}$29$'$00$"$N, 75$^{\circ}$23$'$30$"$W, 2749\,m a.s.l.) to be mainly composed by electromagnetic particles (photons, electrons and positrons) and muons. 


MUSIC-MUSUN codes \cite{kudryavtsev2009, MossEtal2018} and Python/MATLAB scripts calculate the ray tracing, the volcano opacity, the muon energy loss, the emerging muon flux and estimates the period of observation \cite{VesgaRamirez2020, lozano2020, valencia2016}. The code obtains the volcano topography from the SRTM\footnote{\url{https://www2.jpl.nasa.gov/srtm/}} NASA database. The energy loss is calculated considering losses by ionisation and Bremsstrahlung through a standard rock density ($\rho = 2.65$\,g/cm$^3$) or any particular rock compositions in the case of MUSIC-MUSUN codes.

GEANT4 simulations estimates muography affectation by muon forward scattering. We concluded that the muon forward scattering effect is negligible for muon momenta above $5$~GeV/c and incident zenith angles smaller than $85^{\circ}$ from the surface's normal \cite{PeaRodrguez2020}.

We compute the MuTe response by using GEANT4 simulations. This script estimated the deposited energy spectrum of particles impinging the WCD, the photoelectron spectrum of the scintillator strips and the panel attenuation. We found that a vertical muon deposits $\sim240$~MeV in the WCD, losses $\sim2.08$~MeV in the scintillator strip generating an average signal of $40$ photo-electrons. This signal attenuates up to $12\%$ in the panel corner opposite to the SiPM location \cite{VsquezRamrez2020}. Optical parameters (solid angle and acceptance) of MuTe were also computed. MuTe can reconstruct 3481 trajectories with a maximum acceptance of $3.69$~cm$^2$sr \cite{PenaRodriguez2021, VesgaRamirez2020}.

\section{{Instrumentation}}

\begin{figure}[h!]
\centering
\includegraphics[height=1.9in]{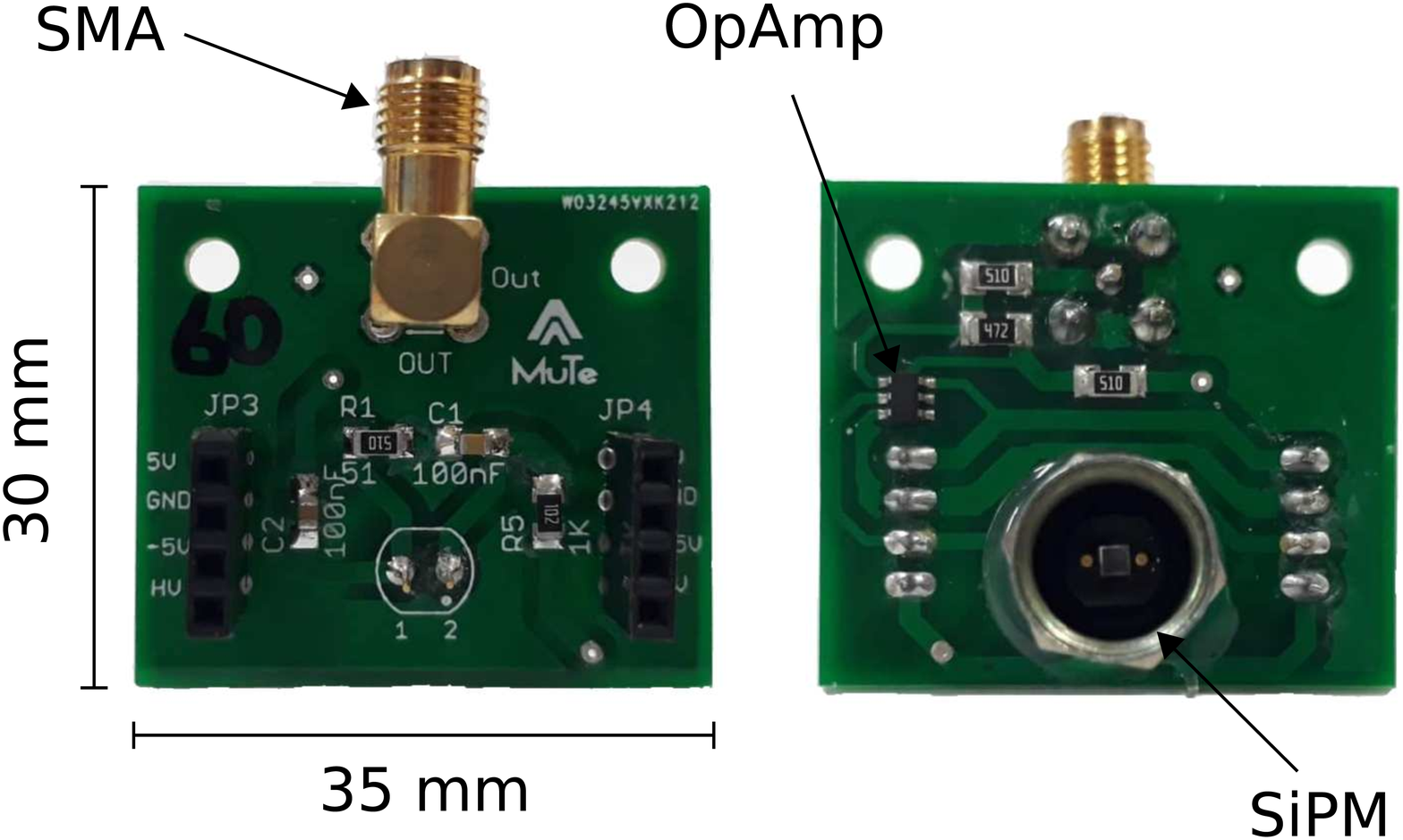}
\includegraphics[height=2.1in]{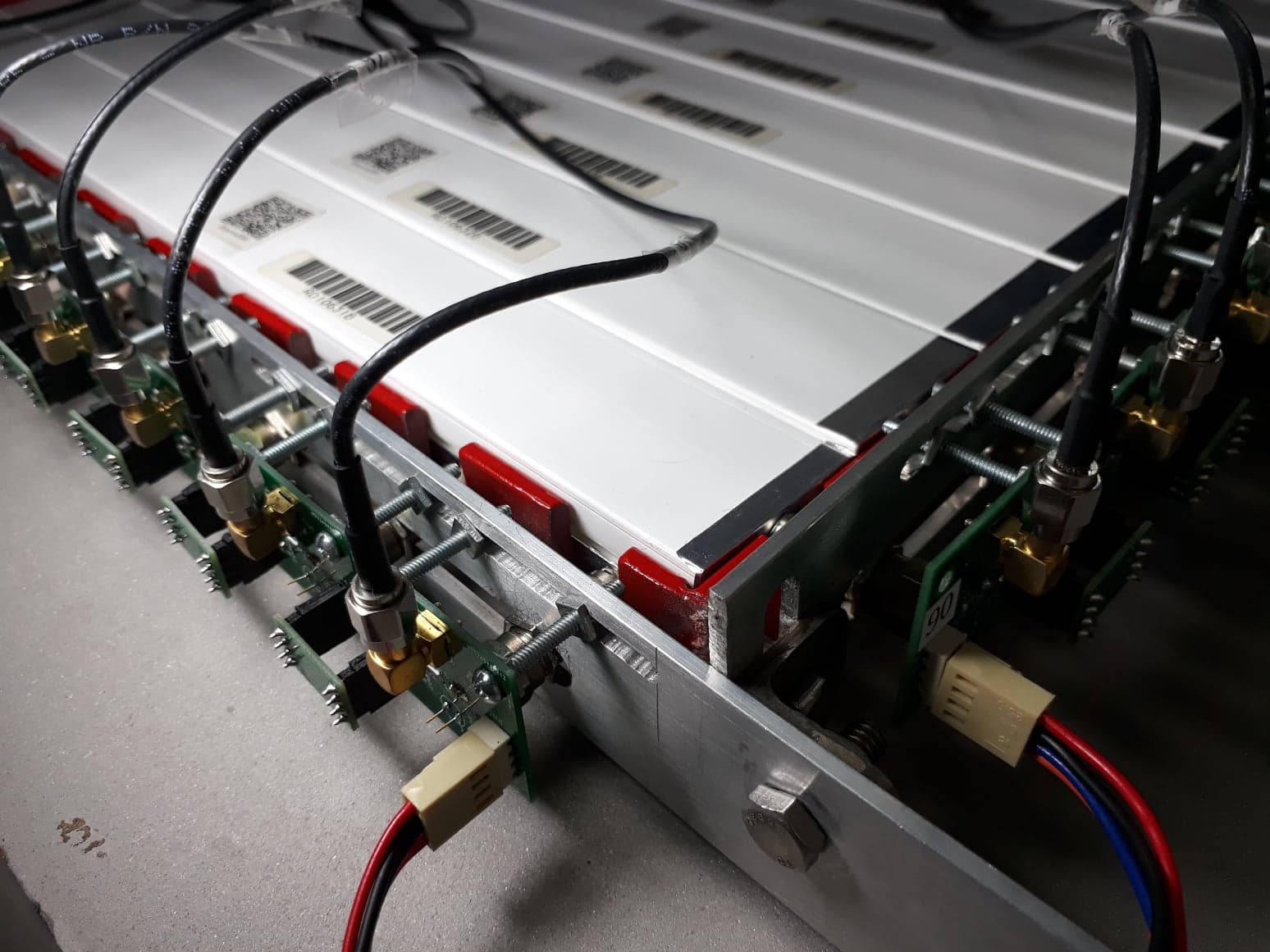}
\caption{(Left) SiPM electronics front-end. (Right) Mechanical assembly of the scintillator bar, the Saint-Gobain BCF-92 WLS fiber, the Hamamatsu S13360-1350CS SiPM, and the transmission cables (coaxial RG-174U).}
\label{f:instrum}
\end{figure}

MuTe combines two detection techniques: a scintillator hodoscope and a water Cherenkov detector. The hodoscope is based on independent modules made of $30$ scintillators bar per panel ($120$~cm x $4$~cm $\times 1$~cm), a WLS fibre (Saint-Gobain BCF-92), a SiPM (Hamamatsu S13360-1350CS) and the electronics frontend. An RG-174U coaxial cable transmits the SiPM pulses to the acquisition system. The SiPM electronics frontend and the scintillator panel assembly are shown in Figure \ref{f:instrum}. 

Two MAROC3A ASICs perform the gain-tuning and discrimination of the $120$~scintillator signals. A couple of Cyclone III FPGAs configure the DAQ slow control parameters. 

We established a discrimination threshold of $8$~photo-electrons considering the SiPM noise analysis. The MuTe controls the SiPMs bias voltage using a programmable power supply (Hamamatsu C11204).

MuTe datasets combine particle flux and environmental data (temperature, barometric pressure, and power consumption) for post-processing, detector status monitoring, and calibration procedures. A GPS pulse-per-second signal synchronises the hodoscope and WCD data.

A PMT (photomultiplier tube Hamamatsu R5912) detects the Cherenkov light from the charged particles crossing the WCD. A high-voltage power supply EMCOC20 biases the PMT from $0$ to $2000$~V. Two 10-bit ADCs digitise the PMT anode and last-dynode signals at a frequency of $40$~MHz. An FPGA Nexys II handles the tasks of thresholding, base-line correction, temporal labelling, and temperature-pressure recording.

The third ADC channel in the WCD digitises a NIM signal from the hodoscope indicating a double coincidence event (hodoscope + WCD). A Cubieboard-2 sets the acquisition parameters (discrimination thresholds and the PMT bias voltage) of the WCD \cite{PeaRodrguez2020, PenaRodriguez2021}.

\begin{figure}[h!]
\centering
\includegraphics[height=2.0in]{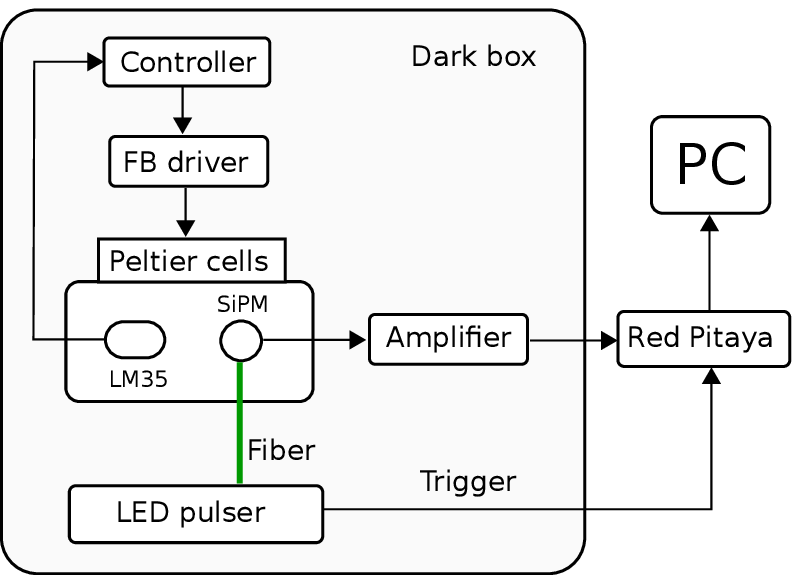}\
\includegraphics[height=2.3in]{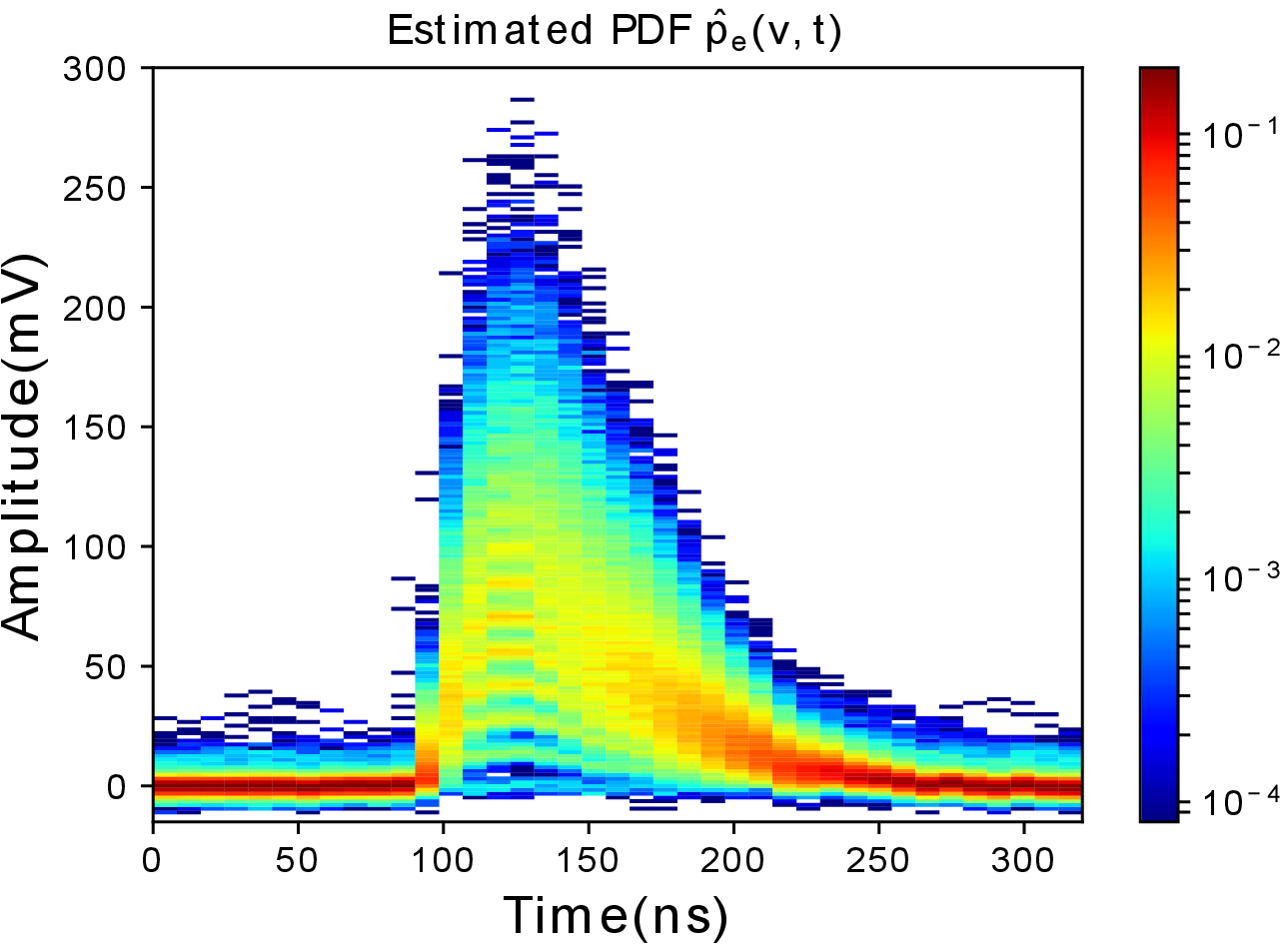}
\caption{(Left) Experimental setup for measuring the gain and noise of the SiPM S13360-1350CS. A $480$~nm pulsed light $10$~ns width at $500$~Hz stimulates the SiPM. The SiPM pulses are digitized at $14$-bit/$125$~MHz. (Right) SiPM pulse waveform quantization.}
\label{f:sipm}
\end{figure}

During the development of the MuTe project, we implemented several specialised equipment and instrumentation for muography. We designed a controlled environment for the calibration of Silicon Photomultipliers (see Figure \ref{f:sipm}). This equipment allows us to control the SiPM temperature, bias voltage, and pulsed-light stimulus. We characterised the SiPM parameters as breakdown voltage, gain, dark count, afterpulsing, and crosstalk, and their dependency on temperature \cite{SanchezVillafrades2021, SanchezVillafrades2018}.

\begin{figure}[h!]
\centering
\includegraphics[height=2.3in]{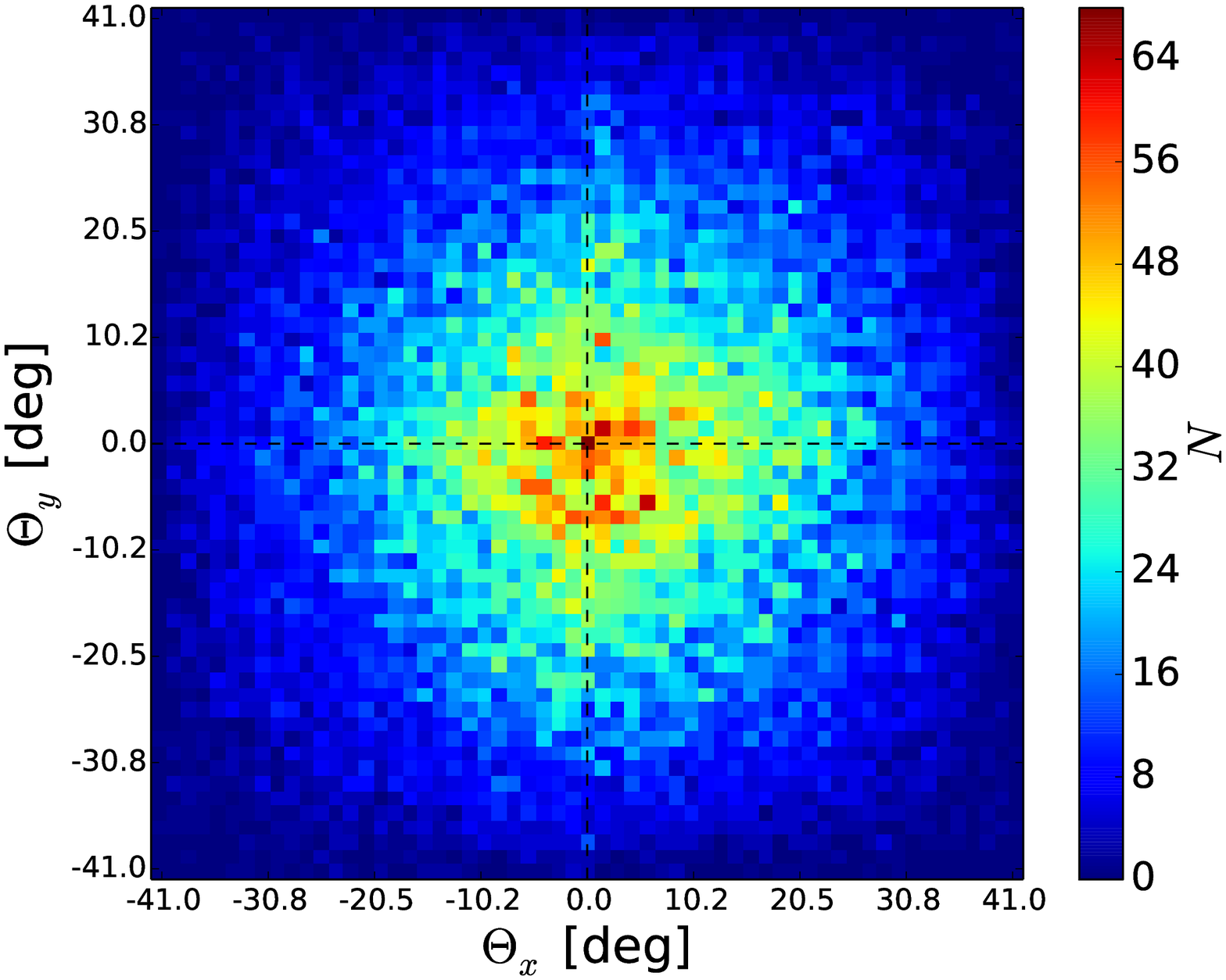}\
\includegraphics[height=2.3in]{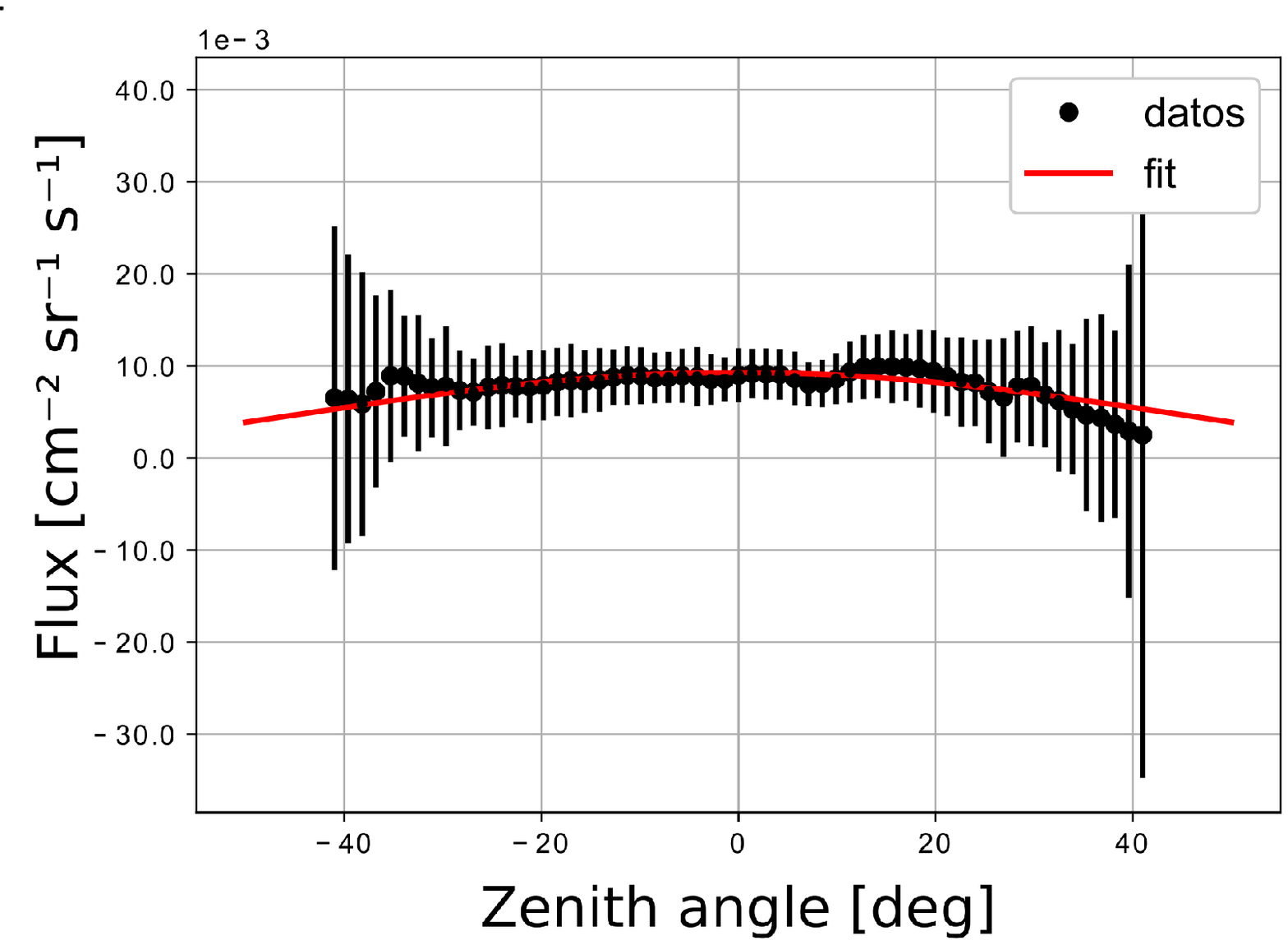}
\caption{(Left) Particle count recorded by the hodoscope during 15 hours with a separation of $13$~cm between panels. (Right) Open sky muon flux profile recorded by the hodoscope. The muon flux measurements (black-line) ranges zenith angles between -40$^{\circ}$ to $40^{\circ}$. A $\cos^n$ function fits the muon flux with $n \sim 2$.}
\label{f:flux}
\end{figure}

We also performed a complete characterisation of the signal attenuation along the scintillator strips \cite{VsquezRamrez2020}. We calibrated the hodoscope response using the atmospheric cosmic ray flux. It allows us also to determine the acceptance confidence window taking into account the variance of the measured open sky muon flux depending on the zenith angle. Figure \ref{f:flux} shows the muon number histogram after 15 hours of recording and the comparison between the estimated (red-line) and measured (black-points) muon flux. The measuring error increases with the zenith angle while the muon flux and the hodoscope acceptance decrease.

\subsection{Mechanics}

We designed the detector mechanics and simulated its structural behaviour under typical environmental conditions through SOLIDWORKS. 
We determined an average temperature of $23^{\circ}$C on the MuTe with maximum temperatures reaching $\sim60^{\circ}$C in the centre of the scintillation panels due to solar radiation. The water volume ($\sim1.7$~m$^3$) of the WCD and the wind flow reduce the telescope temperature.

The mechanical structure of MuTe was found to be robust against tremors and vibrations triggered by volcanic activity. The MuTe vibrations did not exceed 0.04 Hz, guaranteeing its structural integrity \cite{PeaRodrguez2020}.

\section{{ Data analysis}}
\begin{figure}[h!]
\centering
\includegraphics[height=2.2in]{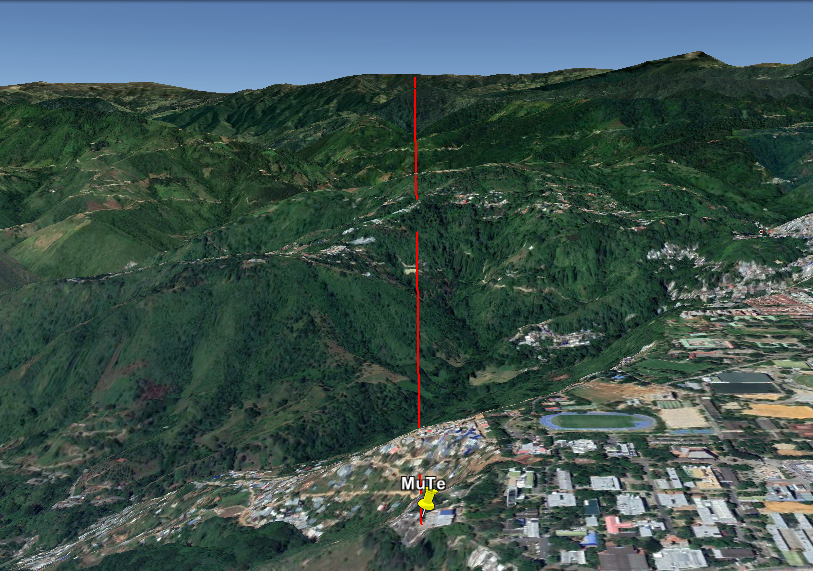}
\includegraphics[height=2.5in]{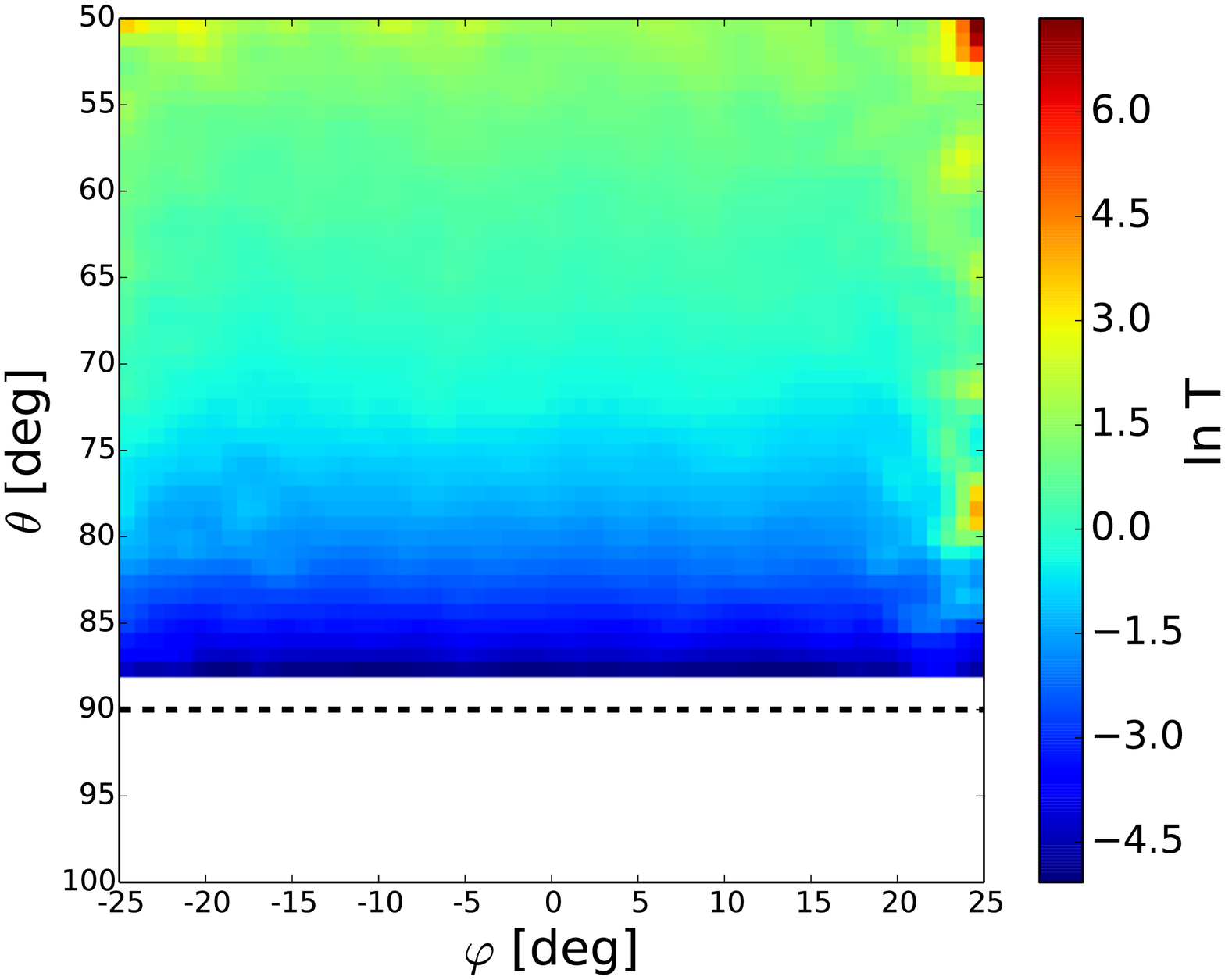}
\caption{(Left) Frontal view of the mountain range scanned by the MuTe. The telescope has an elevation angle of $\sim15^{\circ}$ covering zenith angles from $50^{\circ}$ to $90^{\circ}$ and $10^{\circ}$ under the horizon. (Right panel) Muogram obtained by MuTe after 2-month data recording. The muogram is in terms of the logarithm of transmittance ($T = \ln \Phi/\Phi_0$) with $\Phi$ the emerging muon flux and $\Phi_0$ open sky muon flux. In the open sky zones ($\Theta >$70$^{\circ}$) $T \sim 1$ ($\ln T \sim 0$), while for the mountain region the transmittance decrease ($\ln T < 0$). Below the horizon line, the open sky flux is zero, and the transmittance is infinite.}
\label{f:muogram}
\end{figure}

The MuTe performed its first muogram after two months of data recording pointing towards the mountain range at the northwest of Bucaramanga-Colombia with an elevation of 15$^{\circ}$. The open sky muon flux $\Phi_0$ was estimated considering the data recorded during the hodoscope calibration stage. Figure \ref{f:muogram} shows the MuTe view and the obtained muogram \cite{Nunez2021}.

\subsection{Muography background characterization}

MuTe data analysis allows us to understand the composition of particles impinging the detector during a muography adquisition run. Not only muons impinge the detector, background particles too. We found four background sources in muography: electromagnetic particles of extensive air showers (photons, electrons, and positrons), back-scattering particles, forward scattering muons, and multiple-particle events.

\begin{figure}[h!]
\centering
\includegraphics[height=2.3in]{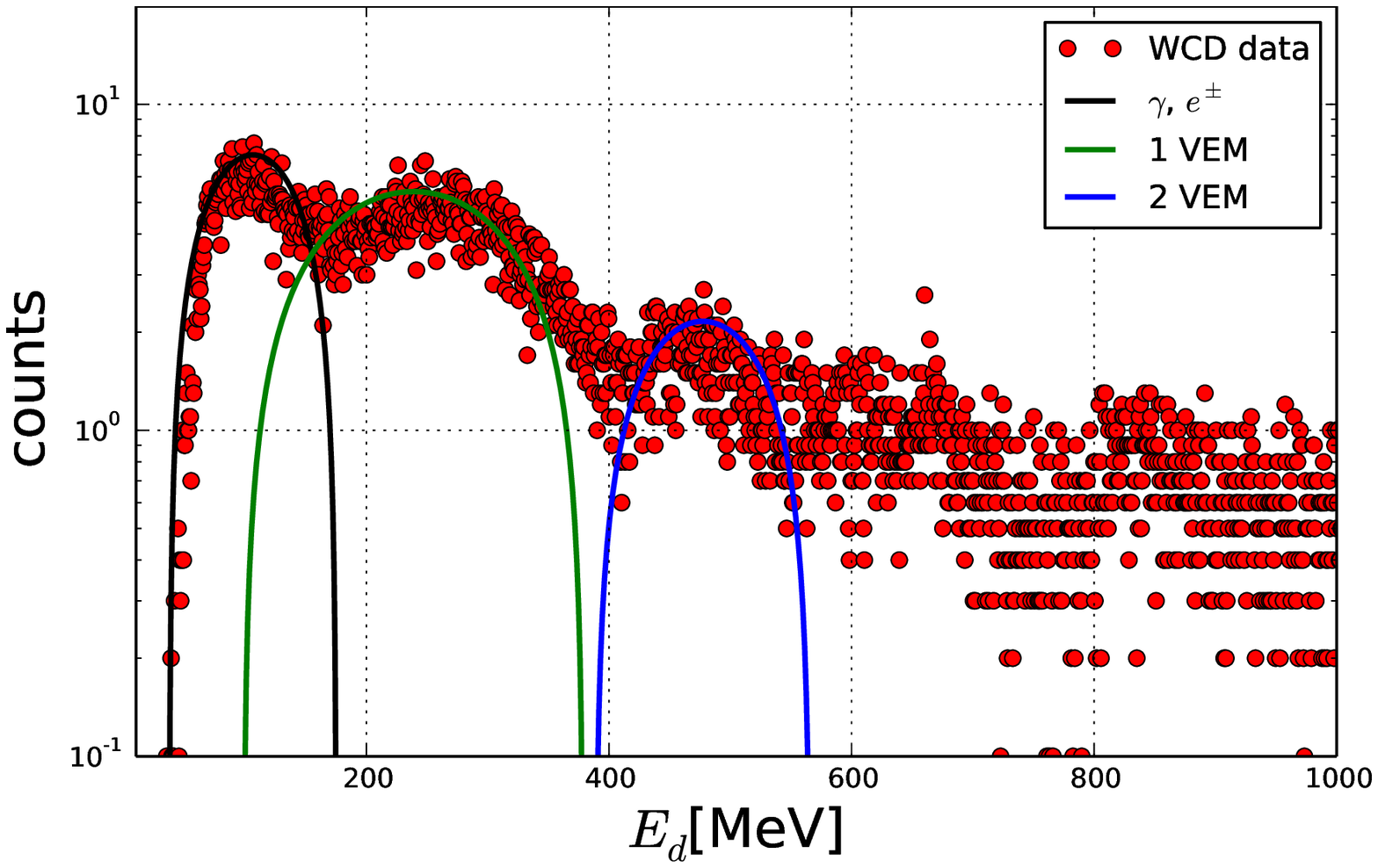}
\includegraphics[height=2.3in]{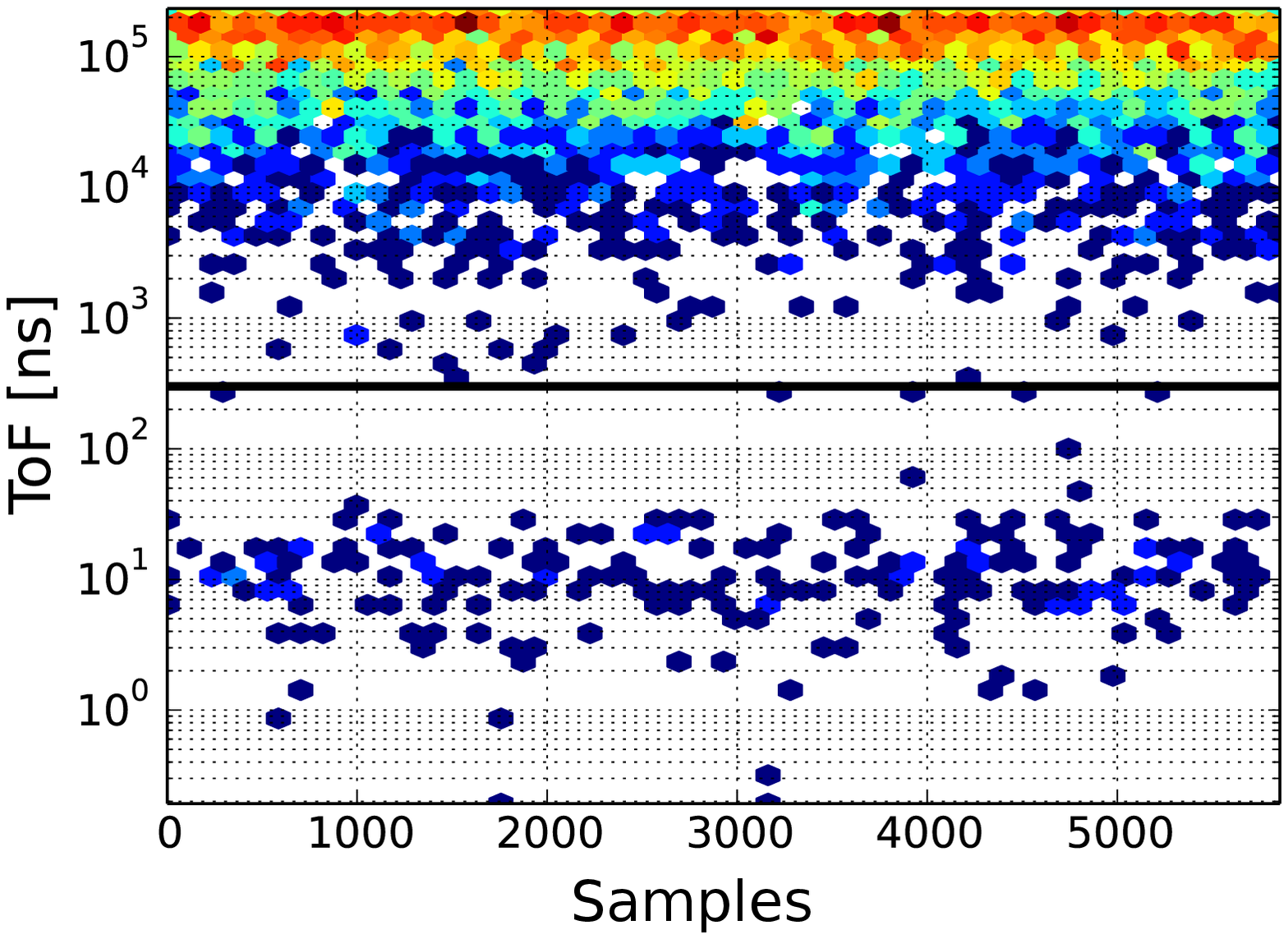}
\caption{(Left) Deposited energy histogram for particles crossing the MuTe during one hour. The black-line hump ($<180$~MeV) corresponds to the electromagnetic component. The green-line hump ($180$~MeV$< Ed <400$~MeV) indicates the deposited energy of muons. Multiple particle events (blue-line) deposit energy above $400$~MeV. (Right-panel) Time-of-Flight of particles traversing the MuTe hodoscope. Single-particle events and correlated background have a ToF $<$100\,ns, while the uncorrelated background has a ToF $>300$~ns.}
\label{f:background}
\end{figure}

We characterised the muography background by analysing the time-of-flight and deposited energy data. Electromagnetic background deposits an energy lower than $180$~MeV and the multiple-particle component deposits an energy above $400$~MeV with a significant contribution at $480$~MeV (equivalent to 2 vertical muons) as shown in Figure \ref{f:background}-(Left).

Multiple-particle component splits in two: correlated particles ($\Delta t < 100$~ns) originated in the same shower and uncorrelated/independent particles ($\Delta t > 300$~ns) as shown in Figure \ref{f:background}-(Right plate). 

The deposited energy information allows us to determine the particle mass. The incident particle momentum was estimated using the ToF measurements, the traversing distance, and the particle mass \cite{PeaRodrguez2021}. We set a threshold to reject muons with momentum below $1$~GeV/c as shown in Figure \ref{f:background}.

We have implemented a machine-learning algorithm to classify and reduce the muography background. Firstly, training a Gaussian Mixture Model to parameterise the deposited energy spectrum and establish thresholds to separate the electromagnetic and multiple-particle background. Then, we use an algorithm to filter low momentum muons ($<1$~GeV/c) by thresholding \cite{villabona2021}.
 
\subsection{Density inversion}

\begin{figure}[h!]
\centering
\includegraphics[height=2.1in]{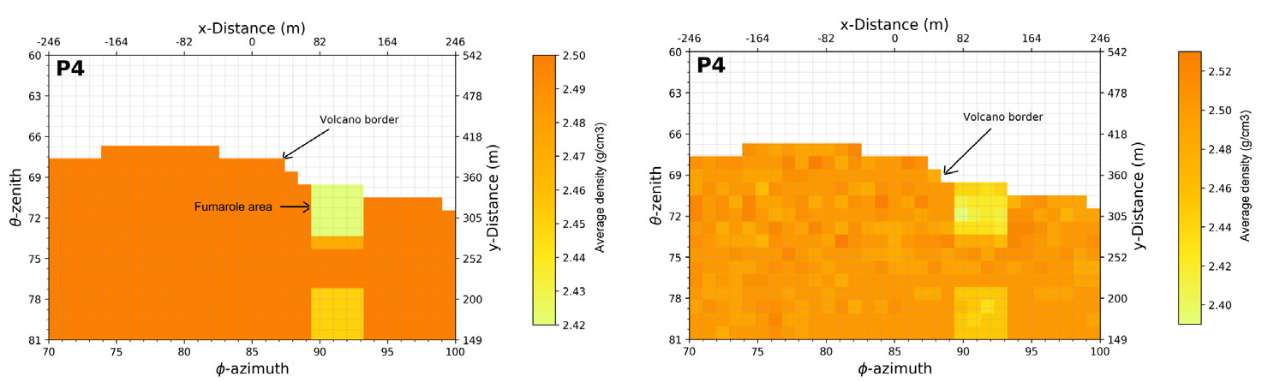}
\caption{(Left) The synthetic density model of the Cerro Machin volcano. (Right) The profile after the inversion. The contrast of densities between the volcanic conduit and the surrounding rock remains \cite{VesgaRamrez2021}.}
\label{f:inversion}
\end{figure}

A crucial step in muography data processing is to solve the inverse problem to determine the inner density distribution of the scanned target. We implemented the Metropolis-Simulated-Annealing algorithm, which inputs an observed muon flux and obtains the best-associated density distribution of the Cerro Machin volcano.

We simulated an inner structure of the Cerro Machín composed by a conduit below the volcano fumaroles zone as shown in Figure \ref{f:inversion}. We improved this model by including rock densities from samples taken from the crater, the dome, and the areas associated with fumaroles. The inversion algorithm reconstructed the density differences inside the emulated Cerro Machín structure from the measured muon flux within a 1$\%$ error. The result showed a remarkable density contrast between the volcanic duct, the encasing rock, and the fumarole area \cite{VesgaRamrez2021}.

\section*{Discussion}

The MuTe project plays a significant role in the muography development in Colombia, Argentina and the Latin American region. We summarise our contributions from creating a simulation framework for muography, instrumentation implementations, and data analysis. 

We simulated the physics behind the muography technique, the cosmic ray flux, the geomagnetic modulation, the creation of extensive air showers in the Earth's atmosphere, the muon flux passing through the geologic structure, the energy loss and scattering of muons, and the detector response. This model can be adapted to any geographical location and detector technology.

We designed a hybrid muon telescope combining a tracking system and a water Cherenkov detector. We can identify muography background sources, electromagnetic particles, back-scattering particles, forward scattering muons, and multiple-particle detection during the data analysis. Finally, we have separated the background components using the deposited energy, tracking, and momentum information. This approach allows us to get extra information about particles impinging the detector.

The MuTe project also addresses the inverse problem of estimating the density distribution of volcanoes. We solved the optimisation problem using a simulated annealing algorithm. We reconstructed the inner structure of the emulated volcano with an error of 1$\%$. The muography program of the Universidad Industrial de Santander offers a training research environment for the new generation of Colombian interdisciplinary particle scientists.

\section*{Acknowledgements}
The authors recognise the financial support of Departamento Administrativo de Ciencia, Tecnología e Innovación de Colombia (ColCiencias) under contract FP44842-082-2015 and to the Programa de Cooperación Nivel II (PCB-II) MINCYT-CONICET-COLCIENCIAS 2015, under project CO/15/02. We are also very thankful to LAGO and the Pierre Auger collaborations for their continuous support.

\bibliographystyle{unsrt}
\bibliography{Muographers.bib}

\begin{thebibliography}{10}

\bibitem{Stern2004}
C.~R. Stern.
\newblock Active andean volcanism: its geologic and tectonic setting.
\newblock {\em Revista geol{\'{o}}gica de Chile}, 31(2), December 2004.

\bibitem{Reath2019}
K.~Reath, M.~Pritchard, M.~Poland, F.~Delgado, S.~Carn, D.~Coppola, B.~Andrews,
  S.~K. Ebmeier, E.~Rumpf, S.~Henderson, S.~Baker, P.~Lundgren, R.~Wright,
  J.~Biggs, T.~Lopez, C.~Wauthier, S.~Moruzzi, A.~Alcott, R.~Wessels,
  J.~Griswold, S.~Ogburn, S.~Loughlin, F.~Meyer, G.~Vaughan, and M.~Bagnardi.
\newblock Thermal, deformation, and degassing remote sensing time series ({CE}
  2000{\textendash}2017) at the 47 most active volcanoes in latin america:
  Implications for volcanic systems.
\newblock {\em Journal of Geophysical Research: Solid Earth}, 124(1):195--218,
  January 2019.

\bibitem{cortes2016}
G.~P. Cort\'es.
\newblock Informe de actividad volc\'anica segmento norte de colombia diciembre
  de 2016.
\newblock Technical report, INGEOMINAS, 2016.

\bibitem{agudelo2016}
A.~Agudelo.
\newblock Informe t\'ecnico de actividad de los volcanes nevado del {Huila,
  Purac\'e y Sotará}, durante el per\'iodo de diciembre de 2016.
\newblock Technical report, Servicio Geol\'ogico Colombiano, 2016.

\bibitem{munoz2017}
E.~Mu{\~n}oz.
\newblock Informe mensual de actividad de los volcanes {Galeras, Cumbal,
  Chiles, Cerro Negro, Las Animas, Do\~na Juana y Azufral}.
\newblock Technical report, INGEOMINAS, 2017.

\bibitem{VesgaRamirez2020}
A.~Vesga-Ram{\'{\i}}rez, D.~Sierra-Porta, J.~Pe{\~n}a-Rodr{\'{\i}}guez, J.~D.
  Sanabria-G{\'o}mez, M.~Valencia-Otero, C.~Sarmiento-Cano,
  M.~Su{\'a}rez-Dura{\'a}n, H.~Asorey, and L.~A. Nu{\~n}ez.
\newblock Muon tomography sites for colombian volcanoes.
\newblock {\em Annals of Geophysics}, 63(6), December 2020.

\bibitem{Murcia2010}
H.~F. Murcia, M.F. Sheridan, J.~L. Mac{\'{\i}}as, and G.~P. Cort{\'{e}}s.
\newblock {TITAN}2d simulations of pyroclastic flows at cerro mach{\'{\i}}n
  volcano, colombia: Hazard implications.
\newblock {\em Journal of South American Earth Sciences}, 29(2):161--170, March
  2010.

\bibitem{DelaCruzReyna2009}
S.~De la~Cruz-Reyna and A.~L. Martin~Del Pozzo.
\newblock The 1982 eruption of el chich{\'{o}}n volcano, mexico: Eyewitness of
  the disaster.
\newblock {\em Geof{\'{\i}}sica Internacional}, 48(1), January 2009.

\bibitem{SierraportaEtal2018}
H.~Asorey, R.~Calder\'on-Ardila, C.R. Carvajal-Bohorquez,
  S.~Hern\'andez-Barajas, L.~Mart\'inez-Ram\'irez, A.~Jaimes-Motta,
  F.~Le\'on-Carre{\~n}o, J.~Pe{\~n}a-Rodr\'iguez, J.~Pisco-Guavabe, J.D.
  Sanabria-G\'omez, M.~Su\'arez-Dur\'an, A.~V\'asquez-Ram\'irez,
  K.~Forero-Guti\'errez, J.~Salamanca-Coy, L.A. N\'u{\~n}ez, and
  D.~Sierra-Porta.
\newblock Astroparticle projects at the eastern colombia region: facilities and
  instrumentation.
\newblock {\em Scientia et technica}, 23(3):391--396, 2018.

\bibitem{PenarodriguezEtal2018B}
H.~Asorey, R.~Calder\'on-Ardila, K.~Forero-Guti\'errez, L.A. N\'u{\~n}ez,
  J.~Pe{\~n}a-Rodr\'iguez, J.~Salamanca-Coy, J.D. Sanabria-G\'omez,
  J.~S\'anchez-Villafrades, and D.~Sierra-Porta.
\newblock minimute: A muon telescope prototype for studying volcanic structures
  with cosmic ray flux.
\newblock {\em Scientia et technica}, 23(3):386--390, 2018.

\bibitem{Asorey2018}
H.~Asorey, L.~A. N{\'{u}}{\~{n}}ez, and M.~Su{\'{a}}rez-Dur{\'{a}}n.
\newblock Preliminary results from the latin american giant observatory space
  weather simulation chain.
\newblock {\em Space Weather}, 16(5):461--475, May 2018.

\bibitem{GrisalescasadiegosSarmientocanoNunez2020}
J.~Grisales-Casadiegos, C.~Sarmiento-Cano, and L.A. N{\'u}{\~n}ez.
\newblock Impact of global data assimilation system atmospheric models on
  astroparticle showers.
\newblock {\em arXiv preprint arXiv:2006.01224}, 2020.

\bibitem{Tapia2017}
A.~Tapia, D.~Due{\~{n}}as, J.~Rodr\'iguez, J.~Betancourt, and
  D.~A.~Mart{\'{\i}}nez Caicedo.
\newblock Preliminary monte carlo simulation study of the structure of the
  galeras volcano using muon tomography.
\newblock In {\em Proceedings of 38th International Conference on High Energy
  Physics {\textemdash} {PoS}({ICHEP}2016)}. Sissa Medialab, February 2017.

\bibitem{Parra2019}
J.~S.~Useche Parra and C.~A.~{\'{A}}vila Bernal.
\newblock Estimation of cosmic-muon flux attenuation by monserrate hill in
  bogota.
\newblock {\em Journal of Instrumentation}, 14(02):P02015--P02015, February
  2019.

\bibitem{VsquezRamrez2020}
A.~V{\'{a}}squez-Ram{\'{\i}}rez, M.~Su{\'{a}}rez-Dur{\'{a}}n, A.~Jaimes-Motta,
  R.~Calder{\'{o}}n-Ardila, J.~Pe{\~{n}}a-Rodr{\'{\i}}guez,
  J.~S{\'{a}}nchez-Villafrades, J.D. Sanabria-G{\'{o}}mez, H.~Asorey, and L.A.
  N{\'{u}}{\~{n}}ez.
\newblock Simulated response of {MuTe}, a hybrid muon telescope.
\newblock {\em Journal of Instrumentation}, 15(08):P08004--P08004, August 2020.

\bibitem{Guerrero2019}
I.~D. Guerrero, D.~F. Cabrera, J.~C. Paz, J.~D. Estrada, C.~A. Villota, E.~A.
  Velasco, F.~E. Fajardo, O.~Rodriguez, J.~Rodriguez, D.~Arturo,
  D.~Due{\~{n}}as, D.~Torres, J.~Ramirez, J.~Revelo, G.~Ortega, D.~Benavides,
  J.~Betancourt, A.~Tapia, and D.~A. Martinez-Caicedo.
\newblock Design and construction of a muon detector prototype for study the
  galeras volcano internal structure.
\newblock {\em Journal of Physics: Conference Series}, 1247(1):012020, June
  2019.

\bibitem{MenchacaRocha2014}
A.~Menchaca-Rocha.
\newblock Using cosmic muons to search for cavities in the pyramid of the sun,
  teotihuacan: preliminary results.
\newblock In {\em Proceedings of 10th Latin American Symposium on Nuclear
  Physics and Applications {\textemdash} {PoS}(X {LASNPA})}. Sissa Medialab,
  October 2014.

\bibitem{VesgaRamrez2021}
A.~Vesga-Ram{\'{\i}}rez, J.D. Sanabria-G{\'{o}}mez, D.~Sierra-Porta,
  L.~Arana-Salinas, H.~Asorey, V.A. Kudryavtsev, R.~Calder{\'{o}}n-Ardila, and
  L.A. N{\'{u}}{\~{n}}ez.
\newblock Simulated annealing for volcano muography.
\newblock {\em Journal of South American Earth Sciences}, 109:103248, August
  2021.

\bibitem{Lesparre2012}
N.~Lesparre, J.~Marteau, Y.~D{\'{e}}clais, D.~Gibert, B.~Carlus, F.~Nicollin,
  and B.~Kergosien.
\newblock Design and operation of a field telescope for cosmic ray geophysical
  tomography.
\newblock {\em Geoscientific Instrumentation, Methods and Data Systems},
  1(1):33--42, April 2012.

\bibitem{PeaRodrguez2020}
J.~Pe{\~{n}}a-Rodr{\'{\i}}guez, J.~Pisco-Guabave, D.~Sierra-Porta,
  M.~Su{\'{a}}rez-Dur{\'{a}}n, M.~Arenas-Fl{\'{o}}rez, L.M.
  P{\'{e}}rez-Archila, J.D. Sanabria-G{\'{o}}mez, H.~Asorey, and L.A.
  N{\'{u}}{\~{n}}ez.
\newblock Design and construction of {MuTe}: a hybrid muon telescope to study
  colombian volcanoes.
\newblock {\em Journal of Instrumentation}, 15(09):P09006--P09006, September
  2020.

\bibitem{PeaRodrguez2019}
J.~Pe{\~{n}}a-Rodr{\'{\i}}guez, A.~V{\'{a}}squez-Ram{\'{\i}}rez, J.~D.
  Sanabria-G{\'{o}}mez, L.~A. Nunez, D.~Sierra-Porta, and H.~Asorey.
\newblock Calibration and first measurements of {MuTe}: a hybrid muon telescope
  for geological structures.
\newblock In {\em Proceedings of 36th International Cosmic Ray Conference
  {\textemdash} {PoS}({ICRC}2019)}. Sissa Medialab, July 2019.

\bibitem{CaldernArdila2020}
R.~Calder{\'{o}}n Ardila, H.~Asorey, and A.~Almela.
\newblock Desarrollo de t{\'{e}}cnicas de muongraf{\'{\i}}a para estudios
  densitom{\'{e}}tricos de objetos de importancia estrat{\'{e}}gica.
\newblock {\em {AJEA}}, (5), October 2020.

\bibitem{PeaRodrguez2021}
J.~Pe{\~{n}}a Rodr{\'{\i}}guez, R.~de{\textquotesingle}Le{\'{o}}n Barrios,
  A.~Ram{\'{\i}}rez-Mu{\~{n}}{\'{o}}z, D.~Villabona-Ardila,
  M.~Su{\'{a}}rez-Dur{\'{a}}n, A.~V{\'{a}}squez-Ram{\'{\i}}rez, H.~Asorey, and
  L.~A. N{\'{u}}{\~{n}}ez.
\newblock Muography background sources: simulation, characterization, and
  machine-learning rejection.
\newblock In {\em Proceedings of 37th International Cosmic Ray Conference
  {\textemdash} {PoS}({ICRC}2021)}. Sissa Medialab, August 2021.

\bibitem{Lesparre2011}
N.~Lesparre, D.~Gibert, and J.~Marteau.
\newblock Bayesian dual inversion of experimental telescope acceptance and
  integrated flux for geophysical muon tomography.
\newblock {\em Geophysical Journal International}, 188(2):490--497, November
  2011.

\bibitem{AsoreyEtal2016a}
H.~Asorey, L.A. N{\'u}{\~n}ez, M.~Su{\'a}rez-Dur{\'a}n, L.A. Torres-Ni{\~n}o,
  M.~Rodr{\'\i}guez-Pascual, A.J. Rubio-Montero, and R.~Mayo-Garc{\'\i}a.
\newblock The latin american giant observatory: a successful collaboration in
  latin america based on cosmic rays and computer science domains.
\newblock In {\em Cluster, Cloud and Grid Computing (CCGrid), 2016 16th
  IEEE/ACM International Symposium on}, pages 707--711. IEEE, 2016.

\bibitem{SarmientocanoEtal2020}
C.~Sarmiento-Cano, M.~Su\'arez-Dur\'an, R.~Calder\'on-Ardila,
  A.~V\'asquez-Ram\'{\i}rez, A.~Jaimes-Motta, S.~Dasso, I.~Sidelnik, L.A.
  N\'u{\~n}ez, and H.~Asorey.
\newblock {Performance of the LAGO water Cherenkov detectors to cosmic ray
  flux}.
\newblock {\em arXiv preprint arXiv:2010.14591}, 10 2020.

\bibitem{valencia2016}
M.~L. Valencia-Otero.
\newblock {Estudio de las componentes de secundarios en cascadas originadas por
  rayos c\'osmicos para aplicaciones sobre estructuras geol\'ogicas}.
\newblock Master's thesis, Universidad Industrial de Santander, 2016.
\newblock Bachelor's Thesis.

\bibitem{kudryavtsev2009}
V.A. Kudryavtsev.
\newblock Muon simulation codes music and musun for underground physics.
\newblock {\em Computer Physics Communications}, 180(3):339--346, 2009.

\bibitem{MossEtal2018}
H.~Moss, A.~Vesga-Ram{\'\i}rez, V.A. Kudryavtsev, L.A. N{\'u}{\~n}ez, and
  D.~Sierra-Porta.
\newblock Muon tomography for the cerro mach{\'\i}n volcano.
\newblock Technical report, Department of Physics \& Astronomy, University of
  Sheffield, Sheffield, United Kingdom, 2018.

\bibitem{lozano2020}
L.~V. Gir\'on-Lozano.
\newblock {C\'alculo de p\'erdida de energ\'ia de muones que interact\'uan con
  diferentes tipos de roca: aplicaci\'on al volc\'an Cerro Mach\'in para el
  proyecto MuTe, en el rango de energ\'ia de 1 a 1000 GeV}.
\newblock Master's thesis, Universidad del Tolima, 2020.

\bibitem{PenaRodriguez2021}
J.~Pe\~na Rodr\'iguez.
\newblock {\em Diseño y calibraci\'on de un telescopio de muones h\'ibrido
  para estudios vulcanol\'ogicos}.
\newblock PhD thesis, Universidad Industrial de Santander, 2021.

\bibitem{SanchezVillafrades2021}
J.~S\'anchez-Villafrades, J.~Pe\~na Rodr\'\i{}guez, H.~Asorey, and L.~A.
  N\'u\~nez.
\newblock {Characterization and on-field performance of the MuTe Silicon
  Photomultipliers}.
\newblock 2 2021.

\bibitem{SanchezVillafrades2018}
J.~S\'anchez-Villafrades, J.~Pe\~na Rodr\'\i{}guez, R.~Calderón-Ardila, and
  L.~A. N\'u\~nez.
\newblock Control de temperatura para el estudio del voltaje de ruptura de
  {SiPMs}.
\newblock In {\em Congreso Internacional de Ciencias Básicas e Ingeniería
  {\textemdash} ({CICI}2018)}. Universidad de los Llanos, August 2018.

\bibitem{Nunez2021}
L.~A. Nunez, R.~de~Leon-Barrios, J.~Pe{\~{n}}a-Rodr{\'{\i}}guez, J.~D.
  Sanabria-G{\'{o}}mez, A.~V{\'{a}}squez-Ram{\'{\i}}rez,
  R.~Calder{\'{o}}n-Ardila, C.~Sarmiento-Cano, A.~Vesga-Ramirez,
  D.~Sierra-Porta, M.~Su{\'{a}}rez-Dur{\'{a}}n, and H.~Asorey.
\newblock Muography for the colombian volcanoes.
\newblock In {\em Proceedings of 37th International Cosmic Ray Conference
  {\textemdash} {PoS}({ICRC}2021)}. Sissa Medialab, August 2021.

\bibitem{villabona2021}
A.~Ram\'irez-Mu\~noz and D.~Villabona-Ardila.
\newblock Discriminaci\'on del ruido de fondo en muograf\'ia usando t\'ecnicas
  de aprendizaje automatizado.
\newblock Master's thesis, Universidad Industrial de Santander, 2021.
\newblock Bachelor's Thesis.

\end{thebibliography}

\end{document}